\documentstyle[aps,prl,floats,epsf]{revtex}
\begin{document}
\advance\textheight by 0.2in
\draft
\twocolumn[\hsize\textwidth\columnwidth\hsize\csname@twocolumnfalse%
\endcsname
\title{Rippling patterns in aggregates of myxobacteria\\
	arise from cell-cell collisions}

\author{Uwe B\"orner$^\ast$,  Andreas Deutsch$^\ast$, Hans Reichenbach$^\dag$
and Markus B\"{a}r$^\ast$}

\address{$^\ast$Max Planck Institute for Physics of Complex Systems,\\  
	N\"othnitzer Stra{\ss}e 38, 01187 Dresden, Germany\\ 
	$^\dagger$ GBF - Gesellschaft f\"ur Biotechnologische Forschung mbH,
	Abteilung Naturstoffchemie,\\
	Matscheroder Weg 1,  38124 Braunschweig, Germany \\ 
}

\date{17 July 2001}
\maketitle 

\begin{abstract}

%
Experiments with myxobacterial aggregates reveal standing waves called
rippling patterns.
Here, these structures are modelled with a simple discrete model 
based on the interplay between migration and collisions of cells.
Head-to-head collisions of cells result in cell reversals.
%
To correctly reproduce the rippling patterns, a refractory phase after each
cell reversal has to be assumed, during which further reversal is prohibited.
%
%
The duration of this phase 
determines the wavelength and period of the ripple patterns 
as well as the reversal frequency of single cells. 
\end{abstract}

]

\section*{}
Pattern formation in aggregates of bacteria and amoebae is 
a widely observed phenomenon\cite{1,2}. 
Complex colonial patterns of spots, stripes and rings are produced
by {\em E. coli}\cite{3&4}. 
{\em Bacillus subtilis} exhibits branching patterns during colonial
growth\cite{5}. 
Two- and three-dimensional circular waves and spirals have been observed
during aggregation of the eukaryotic slime mold {\em Dictyostelium
discoideum} [5-8], often closely resembling patterns in chemical
reactions\cite{12}.
These patterns are typically modelled with continuous reaction-diffusion
equations describing the spatiotemporal evolution of the cell density and
concentrations of chemoattractants and nutrients, {\it e. g.}  in the cases of
{\em Dictyostelium discoideum}\cite{1,13,14} and {\em E. coli}\cite{15,16}.
An alternative approach uses discrete models describing the motion 
of individual cells and has been previously employed to study swarm
behavior [14-16].
%
%
%

The prokaryotic soil bacterium {\em Myxococcus xanthus}\cite{20,21} is one of
the most intriguing examples for morphogenesis and pattern formation.
Like the slime mold {\em Dictyostelium discoideum}, {\em M. xanthus} exhibits
social behaviour and a complex developmental cycle.
As long as there is sufficient food supply, vegetative cells feed on 
other bacterial species, grow and divide. 
But when nutrients run short, bacteria start to aggregate and
finally build a  multicellular structure, the fruiting
body. 
%
%
In order to maintain this life cycle intercellular communication is
essential. 
%
%
Although a phospholipid chemoattractant has been
identified\cite{22} recently, the key role is ascribed to 
interactions occuring 
through cell-cell contact during collisions\cite{20}.
An experimental example for the rippling phenomenon\cite{23,24}
is displayed in Fig. 1.
Bacteria organize into equally spaced ridges (dark regions) that are separated
by regions with low cell density (light regions); for a movie see \cite{27}.
Rippling patterns were first discovered by one of us (H. Reichenbach) and
originally named {\em oscillatory waves}.
%
We examine the temporal dynamics of the density profile along a
one-dimensional cut indicated by the white line in Fig. 1a.
The resulting space-time plot is shown in Fig. 1b and reveals a periodically
oscillating standing wave pattern
%
%
superimposed by spatiotemporal noise.
\begin{figure}
\label{FIG1}
\epsfxsize=65mm
\centerline{\epsffile{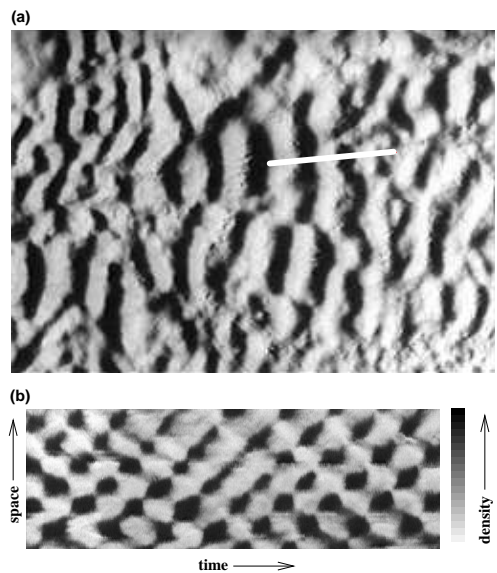}}
\vspace{2mm}
\caption{{\bf (a)} Snapshot from a rippling sequence, taken from a time-lapse
movie. White bar: 0.5 mm. {\bf (b)} Space-time plot of the density profile
along the white line in (a).
}
\end{figure}

In the following, we present a model based on the dynamics of individual cells
for the formation of ripple patterns during the aggregation of myxobacteria.
%
%
%
We will show how certain collision rules between cells on the
``microscopic'' level lead to the observed macroscopic pattern and reproduce
the characteristics of single cell behaviour. 
%
%
%
The basic rules of the model are derived from experimental results by Sager and
Kaiser\cite{25}: 
Within the rippling phase cells are found to move on linear
paths parallel to each other about a distance of one wavelength. 
When two opposite moving cells collide head-on, they reverse their
gliding direction due to exchange of a small, membrane-associated protein
called C-factor. 
Furthermore, the model assumes a
refractory phase in which cells can not respond to the signal. 
This
additional ingredient is necessary for the formation of the rippling pattern 
from a random configuration. 
The characteristic wavelength and period is determined by the duration $\tau$
of the refractory phase.
$\tau$ is the
only adjustable parameter; the experimental data are reproduced with a
refractory period of  five minutes. 
%

%
%
%
%

\begin{figure}
\label{FIG2}
\epsfxsize=65mm
\centerline{\epsffile{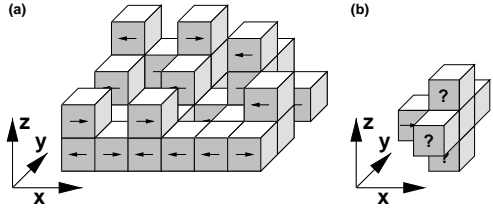}}
\vspace{2mm}
\caption{{\bf (a)} Exemplary configuration of the model, the cell
orientation is indicated by arrows. {\bf (b)} The interaction neighborhood is a
five nodes cross in the $y,z$-plane at that $x$-position the cell is directed
to (here the cell orientation is the $+x$-direction).
}
\end{figure}

In the rippling phase,
cells are densely packed and elongated parallel to each other. 
The organisation of cells into the aligned state can be modeled using
particle-based  models\cite{18,19}. 
Our model is defined on a fixed square grid in the $x-y$-plane and 
assumes discrete space coordinates, analagous to cellular
automaton models. 
The number of cells in the model is conserved and
characterized by the average number $\bar{n}$ of cells per lattice
site; cell death and replication of cells are neglected. 
The discrete $z-$discrete coordinate describes the number of cells 
piled up on top of each other in a given lattice point in the
$x-y$-plane.
The temporal update is done in discrete time steps. 
Cells mainly move along linear paths in the $x$-direction. 
Thus, even a two-dimensional model is a reasonable first step.  
Nevertheless, we also study the three-dimensional case because of 
its experimental relevance. 
The movement of individual bacteria is restricted to  
sheets  with fixed $y$-coordinate.
The coupling in
the $y$-direction is solely due to interaction.
%
A single cell is thus described by a three-dimensional space coordinate
$(x,y,z)$ and an orientation variable
$\phi\in\{-1\mbox{(left)},1\mbox{(right)}\}$ referring to the gliding
direction.
%
Cells interact only via  head-on collisions, {\it
i. e.} cells only sense counterpropagating cells in a certain 
interaction neighborhood. 

The sensitivity of a bacterium to C-factor is described 
by a  {\em clock} variable $\nu$. 
When a sensitive cell collides head-on with other cells, 
(the meaning of collisions will be specified below), it
reverses its gliding direction and is refractory for the next $\tau-1$ time
steps. $\nu$ measures the time since the last reversal, thus a cell with
$\nu<\tau$ is insensitive to C-factor.
%
Overhangs and holes in the bulk are prohibited. 
A bacterial cell is assumed to cover
one node of the lattice, this determines the lattice constants
of 10 $\mu$m in $x$- and 1 $\mu$m in $y$- and $z-$directions 
and the time constant of 1 min.

The temporal update of the model consists of a migration and an
interaction step. 
In the asynchronous migration step cells move according to
their orientation to the neighboring site in $x$-direction. 
If this site is
already occupied, the cell pushes its way between cells of the
adjacent column increasing its height by 1.
With equal probability the cell slips beneath or above the blocking
cell.
This random process causes internal noise.
There is also a diffusion-like noise contribution, because cells are assumed
to rest with small probability $p$ (in the simulations below $p =
0.05$ is used).
%
%
Interaction takes place simultaneously;
every sensitive cell ($\nu\geq\tau$) checks a neighborhood of five nodes 
depending on its orientation $\phi$ (Fig. 2b).
If a cell encounters at least one cell with opposite orientation in this
neighborhood ({\em collision}), the cell
reverses orientation ($\phi\to-\phi$) and will be refractory for $\tau$ time
steps. 
%
The cell is insensitive to neighbors but it can still cause the reversal
of other cells. 
Random initial conditions and periodic boundary conditions are
used throughout.

%


\begin{figure}
\label{FIG3}
\epsfxsize=65mm
\centerline{\epsffile{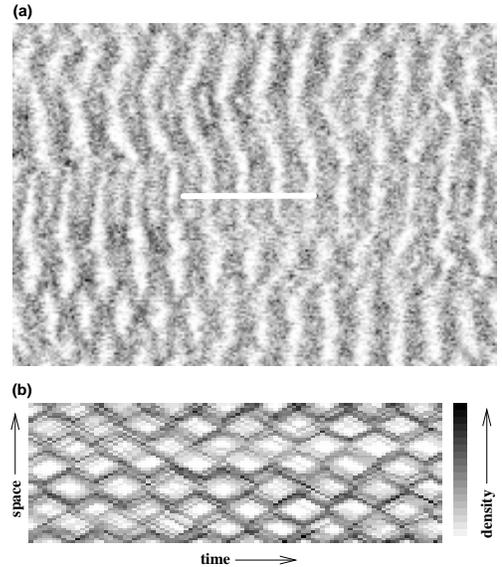}}
\vspace{2mm}
\caption{{\bf (a)} Simulation snapshot in a system of size 220$\times$150
with a refractory phase $\tau=5$ (compare to Fig. 1). Typical pattern emerging
from random initial conditions after ca. 500 time steps. {\bf (b)} Space-time
plot of the density profile along the white line. The gray scale expresses to
the height of the cell column on each site (black corresponds to high
columns).
}
\end{figure}

The model described in the previous section reproduces the
experimentally observed ripple patterns, see Fig. 3. 
%
%
The rippling pattern and the 
temporal evolution obtained in the model (Fig. 3) are in good
agreement with the experimental data of Fig. 1.
Waves propagate equally in both directions, their superposition forms
a standing wave. 
The wavelength of the ripple pattern increases with the duration
$\tau$ of the refractory phase. 
The discrete model enables us to track the single cell behavior.
Typically, cells move over a distance of about half a wavelength before they
reverse their orientation (see Fig. 4c).
%
%
The pattern is easily recognizable - wavelength and period of the ripples have
been reproduced in several indepedent runs and depend only weakly on the
number of cells in the aggregate (variations of the average number $\bar{n}$ of
cells per lattice point between 2 and 10 do not produce significant changes,
results presented here are for $\bar{n}=3$).
Movies of the simulated rippling patterns can be found in \cite{27}.

More insight is obtained by deriving a mean-field theory of the 
discrete model in 2d. 
Such a description uses a hierarchy of rate equations in discrete time
and space, which replace
the discrete state variables by their average numbers. 
The mean-field scheme \cite{26} leads to the following set 
 of $2\tau$ equations:
\begin{eqnarray}
r_1(x,t+1) &=& r_1(x-1,t)-f_r(x-1,t)+r_{\tau}(x-1,t)\nonumber\\
r_2(x,t+1) &=& f_l(x-1,t)\nonumber\\
r_3(x,t+1) &=& r_2(x-1,t)\nonumber\\
&\vdots& \nonumber\\
r_{\tau}(x,t+1) &=& r_{\tau-1}(x-1,t)\nonumber\\[0.5cm]
l_1(x,t+1) &=& l_1(x+1,t)-f_l(x+1,t)+l_{\tau-1}(x+1,t)\nonumber\\
l_2(x,t+1) &=& f_r(x+1,t)\nonumber\\
l_3(x,t+1) &=& l_2(x+1,t)\nonumber\\
&\vdots& \nonumber\\
l_{\tau}(x,t+1) &=& l_{\tau-1}(x+1,t), 
\label{rateq}
\end{eqnarray}
where $r_1$ {\it resp.} $l_1$ are right- {\it resp.} left moving 
cells that can reserve, while $r_2,...,r_{\tau}$  {\it resp.}
$l_2,...,l_{\tau}$ denote right- {\it resp.} left moving cells 
in the various stages of the refractory phase. 
The functions $f_r$ and $f_l$ describe the average numbers of reversals of
  right- {\it resp.} left moving cells. 
The actual form of the reversal function is quite complicated. 
Since the number of particles on one site is still rather small, 
it is not sufficient to use their mean values in the reversal function
as would be the standard approach for rate equations describing 
chemical reactions. 
Instead, one has to specify the distribution of the quantities $l_1,
r_1$ around their mean values and sum over all possible states. 
We have used a Poissonian distribution for this purpose and performed 
a linear stability analysis of the homogeneous stationary state 
$r_1(x,t) = l_1(x,t) = \rho_S $ and $r_i(x,t) = l_i(x,t) = \rho_R$ for
$ i = 2,....,\tau$. 
This state describes a flat layer of cells with equal amounts of 
left- and right-moving bacteria. 
The actual values of $\rho_S$ and $\rho_R$ depend on the parameter
$\tau$ and $\bar{n}$ and should obey $\rho_R = f_r(x,t) = f_l(x,t)$ and 
$\bar{n} = 2 (\rho_S + (\tau - 1) \rho_R )$. 
For $\tau = 5$ and $ \bar{n} = 3$, we find, for example,  $\rho_S 
\approx 0.635$ and $\rho_R \approx 0.216$.  
A detailed derivation of the mean field theory will be published
elsewhere \cite{26}.

The linear stability  analysis  of the rate equations
(\ref{rateq}) with $\bar{n} = 3$, 
reveals a linear instability of this flat layer state against 
an oscillatory instability with finite wavenumber  
for $\tau \geq 4$ min  \cite{26}. 
%
%
%
Moreover, we obtain the wavenumber $k$ with the 
fastest growth rate for $\tau \geq 4 $ min {\it resp.} weakest damping 
for $\tau < 4$ min and the associated frequency $\omega$
as a function of the refractory time $\tau$. 
A comparison of the corresponding wavenumber $\lambda = 2 \pi / k$ and 
period $T = 2 \pi / \omega$ with equivalent quantities 
 extracted from a Fourier analysis
of simulation data of the discrete model  
shows good agreement below and near the threshold $\tau \leq 4$ min, 
see Figs. 4a,b. 
Above the threshold nonlinear effects lead to a deviation of the
predictions from linear stability analysis. 
It is remarkable that the mode with weakest damping 
can be observed directly for $\tau \leq 3$ min.   
It indicates, that the intrinsic noise of the discrete model 
 drives the system out of the linear stable regime and excites 
the modes with weakest damping. 

The wavelength of the pattern in the experiment is about 180 $\mu m$ 
corresponding to 18 cell lengths. The temporal period in the experiment
is found to be around 10 min. 
Thus, a refractory time $\tau \approx 5$ min in the model yields the correct
experimental values for the wavelength as well as for the period.
As a third quantity, we can measure the average reversal
frequency of the individual cells in the simulations taking advantage 
of the discrete, particle-based nature of our model. 
A typical trajectory of an individual cell in the model is displayed
in  Fig. 4c. 
Most of the time the cells in the model ride with  the ripple
crest and get reflected when two crests collide.  
In other words, while the crest form is seemingly unchanged, most 
of the cells that originally constituted the crest 
are now part of a crest propagating in the other direction. 
Occasionally a cell ,,tunnels'' through and continues a longer way 
with the same crest.  
For the refractory time  $\tau = 5$ min, 
the reversal frequency for a single cell in the model 
is about 0.15 reversals per cell and minute in the three-dimensional model and 0.1
reversals per cell and minute in two dimensions (see Fig. 4d) which is 
in the range of  the  experimentally observed
 frequency of 0.081 reversals per cell and minute\cite{25}. 
%
%

While previous experiments have not provided direct information about the
duration of a refractory phase, recent measurements of reversal rates 
of myxobacteria exposed to high concentration of isolated 
C-factor may give a first clue. 
Sager and Kaiser report an increase of the reversal rate by a factor
of 3 compared to normal aggregates and an absolute reversal rate of 
roughly 0.3 reversals per cell per minute\cite{25}. This suggests a refractory
phase between 3 and 4 minutes. 
The reversal rates from the model with a refractory phase
of 5 minutes would increase by a factor of 1.8 for the two-dimensional
and by a factor of 1.2 for the three-dimensional model. 
This small discrepancy between model and experiment may indicate that the 
refractory time depends on the amount of C-factor and decreases 
at high concentrations.

We presented a model for the formation of ripple patterns
during the aggregation of myxobacteria.
The reversal mechanism of cells following collisions  
has to be supplemented by a refractory phase, that specifies a minimum
time between subsequent reversals. 
The duration of this phase determines the wavelength and the
period of the ripple pattern. 
The ,,microscopic'' single cell behavior agrees well with the
experiments on the reversal frequency of cells. 
%
%
Our study strongly suggests experiments with single cells to verify
the refractory hypothesis and to elucidate its biochemical basis. 
Moreover, myxobacterial rippling provides the first example of a new
mechanism for pattern formation, namely one mediated by migration and direct
cell-cell interaction, that may be involved in selforganization processes 
in other multicellular systems.  

\begin{figure}
\label{FIG4}
\epsfxsize=85mm
\centerline{\epsffile{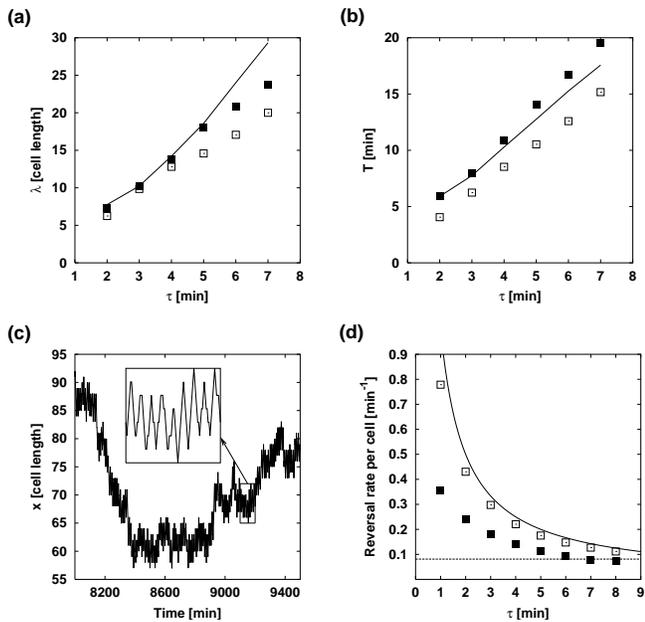}}
\vspace{2mm}
\caption{Ripple wavelength $\lambda$ {\bf (a)} and period T {\bf (b)} versus refractory time
$\tau$ for 2d simulations (solid squares), 3d simulations (open squares) and 2d
mean field theory (solid line). {\bf (c)} Single cell track with a
blow-up of the marked region. {\bf d} Reversal frequency against refractory
time in 2d simulations (solid squares) and 3d simulations (open squares)
compared to experiment (dotted line) and largest possible value
$r_{\mbox{max}}$ (solid line).
}
\end{figure}

\noindent

\noindent
Correspondence and requests for materials should be addressed to M.B.
(e-mail: {\small baer@mpipks-dresden.mpg.de}).

\end{document}